\documentstyle[prd,aps,psfig]{revtex}
\draft
\newcommand{\be}{\begin{equation}}
\newcommand{\ee}{\end{equation}}
\newcommand{\bn}{\begin{eqnarray}}
\newcommand{\en}{\end{eqnarray}}
\newcommand{\bd}{\begin{displaymath}}
\newcommand{\ed}{\end{displaymath}}
\newcommand{\bnn}{\begin{eqnarray*}}
\newcommand{\enn}{\end{eqnarray*}}
\newcommand{\n}{{\not\hspace{-0.5ex}\nabla}}
\newcommand{\N}{{\not\hspace{-0.8ex}D_L}}
\begin{document}
\title{Vacuum Polarization of Massless Spinor Field in Global
Monopole Spacetime} 
\author{ E.R. Bezerra de Mello \thanks{e-mail:
emello@fisica.ufpb.br}, V.B. Bezerra \thanks{e-mail:
valdir@fisica.ufpb.br}, and N.R. Khusnutdinov \thanks{On leave from
Kazan State Pedagogical University, Kazan, Russia; e-mail:
nail@dtp.ksu.ras.ru}}
\address{ Departamento de F\'{\i}sica, Universidade Federal da
Para\'{\i}ba, \\ 
Caixa Postal 5008, CEP 58051-970 Jo\~ao Pessoa, Pb, Brazil}
\date{\today}
\maketitle 
\begin{abstract} 
We calculate the renormalized vacuum average of the
energy-momentum tensor of massless left-handed spinor field in
the pointlike global monopole spacetime using point-separation
approach. The general structure of the vacuum average of the
energy-momentum tensor is obtained and expressed in terms 
of $<T^0_0>^{ren}$ component, explicit form of which is analyzed
in great details for arbitrary solid angle deficit. 
\end{abstract}
\pacs{98.80.Cq, 14.80.Hv, 95.30.S}
\section{Introduction}
It is well-known that different types of topological objects may
have been created by vacuum phase transition in the early
Universe \cite{Kibble,Vilenkin}. These include domain walls,
cosmic strings and monopoles. Among them, cosmic strings and
monopoles seem to be the best candidates to be observed. 

Global monopole is a heavy object formed in the phase transition
of a system composed of a self-coupling scalar field triplet
$\phi^a$ whose original global $O(3)$ symmetry is spontaneously
broken to $U(1)$. The scalar matter field plays the role of an
order parameter which outside the monopole core acquires a
nonvanishing value. The main part of the monopole's energy is
concentrated into its small core. The simplest model which gives
rise to a global monopole is described by the Langrangian
density below, and was analysed by Barriola and Vilenkin
\cite{BarriolaVilenkin} 
\be
L= {1\over 2} (\partial_\mu \phi^a)( \partial^\mu \phi^a ) - {1 \over
4} \lambda (\phi^a \phi^a - \eta^2)^2\ ,
\ee
Coupling this matter field with the Einstein equations, a
spherically symmetric metric tensor given by the line element 
\be
ds^2 = -B(r)dt^2 + A(r)dr^2 + r^2(d\theta^2 + \sin^2\theta
d\varphi^2)\ ,
\ee
presents solutions for the functions $B(r)$ and $A(r)$ far from
the monopole's core given by
\be
B = A^{-1}= 1 - 8\pi \eta^2 - 2 M/r\ .
\ee
The mass parameter $M\sim M_{core}$. Numerical details concerning
with this function can be seen in Ref. \cite{HarariLousto}.  
Neglecting the mass term we get the pointlike global
monopole spacetime with metric
\be
ds^2 = -\alpha^2 dt^2 + dr^2/\alpha^2 + r^2(d\theta^2 +
\sin^2\theta d\varphi^2)\ ,
\label{Metric}
\ee
where the parameter $\alpha^2 = 1 - 8\pi \eta^2$. The energy momentum
tensor of this monopole has diagonal form and reads: $T^t_t =
T^r_r = (\alpha^2 - 1)/r^2$ and $T^\theta_\theta =
T^\varphi_\varphi = 0$. The main peculiarity of this space is
a solid angle deficit which is the difference between the solid angle
in the flat spacetime $4\pi$ and the solid angle in the  
global monopole spacetime which is given by $4\pi\alpha^2$. For
$\alpha < 1$ one has solid angle deficit and for $\alpha >1$ one
has the solid angle excess. (We would like to call attention
that the physical values for $\alpha$ predicted by field theory is
smaller than unit\footnote{In fact for a typical grand
unified theory the parameter $\eta$ is of order $10^{16}Gev$. So
$1 - \alpha^2 = 8\pi \eta^2 \sim 10^{-5}$}). The spacetime
produced by a global monopole has no Newtonian gravitational
potential in spite of the geometry produced by this heavy object
has a non-vanishing curvature. For this reason the mass of the 
monopole is divergent and proportional to the distance from monopole
origin \cite{BarriolaVilenkin}. In context of the monopole
formation the cosmological horizon is a natural cutoff distance
for the monopole's mass. 

Although the global monopole has no Newtonian
gravitational potential it gives enormous tidal acceleration
$a\sim 1/r^2$ which is important from the cosmological point of 
view and may be used for obtaining upper bound on the number
density of them in the Universe, which is at most one
global monopole in the local group galaxies \cite{HiscockPRL}.
However, the numerical simulations made by Bennet and Rhie show that
real upper boundary is smaller than that given in \cite{HiscockPRL}
by many orders \cite{BennetRhie}. In fact, one has scaling
solution with a few global monopoles per horizon volume
\cite{BennetRhie}.  

The quantum effects due to the monopole background in the matter
fields have been considered explicitly for scalar field in Ref. 
\cite{MazitelliLousto} and by general consideration in
Ref.\cite{Hiscock}. It has been shown from general consideration
that the vacuum expectation value of the energy-momentum tensor
of massless fields has the following general form 
\be
<T^\nu_\mu>^{ren} = {S^\nu_\mu(\mu r) \over r^4}\ ,
\ee
where the tensor $S^\nu_\mu$ depends on the arbitrary mass scale
parameter $\mu$ and the metric coefficient $\alpha$. In Ref.
\cite{Hiscock}, it has been assumed that the tensor $S^\nu_\mu$
is the function of the metric parameter $\alpha$, only. Manifest
calculations in scalar case have shown \cite{MazitelliLousto}
that this tensor depends on renormalization mass parameter $\mu$
and it has the following structure $S^\nu_\mu = A^\nu_\mu +
B^\nu_\mu \ln \mu r$, where tensors $A^\nu_\mu$ and $B^\nu_\mu$
depend only on the $\alpha$. This is in agreement with Wald
\cite{WaldPRD} who noted that an unambiguous prescripton for
$<T^\nu_\mu>^{ren}$ cannot be defined without introducing a
length scale. Nevertheless, one-loop Einstein equations do not
depend on scale parameter $\mu$ due to renormalization group
equation.  Back reaction problem in the scalar massless case has
been investigated by Mazzitelli and Lousto
\cite{MazitelliLousto}. It must be noted that only the general
structure of vacuum expectation value of the energy-momentum
tensor has been considered. There, it was not obtained an
explicit form of the tensor $S^\nu_\mu$.  

In this paper we would like to obtain the explicit value of this
tensor, considering the massless spinor field on
the background of the pointlike global monopole with metric
given by Eq.(\ref{Metric}). As opposed to the massless scalar
field case \cite{MazitelliLousto} we get a simpler
expression for the Green function and the energy-momentum tensor
which is obtained in closed form, for arbitrary angle deficit. 

The analysis of the quantum behavior of two massless left-handed
$SU(2)$ doublets fermionic field in the background of a
pointlike monopole, taking into account the magnetic field, has
been developed by Ren \cite{Ren}. There, the Rubakov and Callan
effect was analysed, and found that there appears a small
correction due to the parameter $\alpha$ into the fermion number
condensate. 

This paper is organized as follows. In Sec. \ref{SGF} we
consider the Green functions for massive and massless spinor
fields on the background of the pointlike global monopole and
obtain, in closed form, the Euclidean Green function for
massless left-handed spinor field. In Sec. \ref{EMT} we analyze
this function at coincidence limit and extract all divergencies
from it in manifest form. We obtain also the general structure
of renormalized energy-momentum tensor. Each component of this
tensor may be expressed in terms of only the zero-zero component
$<T^0_0>^{ren}$ which is analyzed with great details. In Sec.
\ref{Conclusion} we summarize our results. Appendices A and
B contain some technical formulas.  The signature of the
spacetime, the sign of Riemann and Ricci tensors are the same as
in Christensen paper \cite{Christensen78}. We use units $\hbar =
c= G = 1$.  
\section{Spinor Green Functions}\label{SGF}
In this section we want to obtain the expression for the fermion
propagator of massive spinor field in the pointlike monopole
spacetime. Massless field will be recognized as particular 
case. The spinor Feynman propagator is defined as follows
\cite{BirrellDevies} 
\be
i{\cal S_F}(x,x') = <0|T(\Psi (x)\overline{\Psi} (x'))|0>\ ,
\ee
where $\overline{\Psi} = \Psi^+\gamma^0$ which, under the
Lorentz transformation $\Psi \to S(\triangle )\Psi$, transforms as
$\overline{\Psi} \to \overline{\Psi}S^{-1}(\triangle )$, where 
$\triangle$ is the parameter of the transformation, and
$S(\triangle )$ is a local representation of the Lorentz group. 

This propagator obeys the following differential equation 
\be
(i\n - M){\cal S_F}(x,x')= {1 \over \sqrt{-g}} \delta^{(4)} (x -
x')I_4\ ,
\label{Green}
\ee
where $g={\rm det}(g_{\mu\nu})$. The covariant derivative
operator in the above equation is 
\be
\n = e^\mu_{(a)}\gamma^{(a)} (\partial_\mu + \Gamma_\mu)\ ,
\ee
$e^\mu_{(a)}$ being the vierbein satisfying the condition 
\be
e^\mu_{(a)}e^\nu_{(b)}\eta^{ab} = g^{\mu\nu}\ ,
\ee
and $\Gamma_\mu$ is the spin connection, given in terms of the
flat spacetime $\gamma$-matrix by 
\be
\Gamma_\mu = -{1\over 4}
\gamma^{(a)}\gamma^{(b)}e^\nu_{(a)}e_{(b)\nu ;\mu}\ .
\ee
The Green function given in (\ref{Green}) is a bispinor, i.e. it
transforms as $\Psi$ at $x$ and as $\overline{\Psi}$ at $x'$. 

If a bispinor ${\cal D_F}(x,x')$ satisfies the differential
equation below
\be
(\Box - M^2 - {1\over 4}R){\cal D_F}(x,x') = -{1\over
\sqrt{-g}} \delta^{(4)} (x-x')I_4\ , \label{GF}
\ee
where the generalized d'Alembertian is expressed by 
\be
\Box = g^{\mu\nu} \nabla_\mu\nabla_\nu = g^{\mu\nu}
(\partial_\mu \nabla_\nu + \Gamma_\mu \nabla_\nu -
\Gamma^\alpha_{\mu\nu} \nabla_\alpha)\ ,
\ee
then the spinor Feynman propagator may be read as 
\be
{\cal S_F}(x,x') = (i\n + M){\cal D_F}(x,x')\ ,
\ee
which shows that the non-minimal coupling to the curvature does play
a role when spinor fields are considered. 

Now after this brief review about the calculation of spinor
Feynman propagator in the general manifold, let us specialize it
to the spacetime of a global monopole. We shall choose the
following basis tetrad:
\be
e^\mu_{(a)} = \left( \begin{array}{cccc}
  1/\alpha & 0 & 0 & 0 \\
  0 & \alpha\sin\theta \cos\varphi & \cos \theta
  \cos\varphi /r & -\sin\varphi /r\sin\theta \\ 
0 & \alpha\sin\theta \sin\varphi & \cos \theta
\sin\varphi /r & \cos\varphi /r\sin\theta \\ 
0 & \alpha\cos\theta & -\sin \theta /r & 0 
                      \end{array}
               \right) \ .
\label{Tetrada}
\ee
For this case, the only non-zero spin connections are
\bn
\Gamma_2 &=& {1-\alpha \over 2} [\gamma^{(1)} \gamma^{(2)}\cos
\varphi + \gamma^{(2)}\gamma^{(3)} \sin \varphi]\ , \\
\Gamma_3 &=& -{1 - \alpha \over 2} [\gamma^{(1)} \gamma^{(2)} \sin
\theta + \gamma^{(1)} \gamma^{(3)} \cos\theta \sin \varphi -
\gamma^{(2)} \gamma^{(3)} \cos\theta \cos\varphi] \sin\theta\ .
\nonumber 
\en
If we had made another choice for the vierbein, for example
$e^\mu_{(a)} = {\rm diag}(1/\alpha , \alpha, 1/r,
1/r\sin\theta)$, the only non-zero spin connection would be $\Gamma_2
= -\alpha \gamma^{(1)} \gamma^{(2)}/2$ and $\Gamma_3 =
-\alpha[\gamma^{(1)} \gamma^{(3)}\sin\theta + \gamma^{(2)}
\gamma^{(3)} \cos\theta ]/2$. Although this tetradic basis is
simpler than previous one given in (\ref{Tetrada}), the spin
connection obtained by the former does not vanish when we take
Minkowski limit $\alpha =1$.  

In order to obtain the explicit form for the differential
equation for the bispinor ${\cal D_F}$ given in
(\ref{GF}) in this geometry we shall adopt the following
representation for the $\gamma$-matrix: 
\be
\gamma^{(0)}=\left( 
\begin{array}{cc}
1&0 \\
0&-1 
\end{array} \right) \ ,
\gamma^{(k)}=\left( 
\begin{array}{cc}
0&\sigma^k \\
-\sigma^k&0 
\end{array} \right) \ ,
\ee
$\sigma^k$ being the $2\times 2$ Pauli matrices. These matrices
above obey the anticommutator relations $\{\gamma^{(a)},
\gamma^{(b)}\} = -2\eta^{ab}$. 

After some intermediate steps, we get the following expression
for the d'Alembertian operator 
\be
\Box = -{1 \over \alpha^2} \partial^2_t + {\alpha^2 \over r^2}
\partial_r (r^2 \partial_r )- {1\over r^2} \vec{L}^2 - {(1 -
\alpha)^2 \over 2r^2} -{1-\alpha \over r^2} \vec{\Sigma} \cdot 
\vec{L} \ ,
\ee
where 
\be
\vec{\Sigma} = \left(\begin{array}{cc}
\vec{\sigma} & 0 \\
0&\vec{\sigma} 
                    \end{array}
                     \right)\ .
\ee
From the above operator we can see that, although it is a
$4\times 4$ matrix differential one, it is diagonal in
block of $2\times 2$ matrices, which means that the two upper
components of Dirac spinor interact with the gravitational field
in the similar way as the two lower components. 

The complete differential operator in Eq. (\ref{GF}) is 
\be
{\cal L} =-{1 \over \alpha^2} \partial^2_t + {\alpha^2 \over r^2}
\partial_r r^2 \partial_r - {1\over r^2} \vec{L}^2 -{1-\alpha
\over r^2}(1 + \vec{\Sigma} \cdot \vec{L}) -M^2\ . 
\ee
Now we shall consider the case where the fermionic field has no
mass. In this case we are able to obtain a closed expression for
the fermionic propagator as follows.

The system which we shall consider consists of a massless
left-handed fermionic field in a global monopole manifold. The
Dirac equation reduces to a $2\times 2$ matrix differential one,
as shown below:
\be
i\N \chi = 0\ ,
\ee 
where
\be
\N = i\left[ {1 \over \alpha} \partial_t - \alpha \sigma^{(r)}
\partial_r - {1\over r} \sigma^{(\theta )}\partial_\theta -{1
\over r\sin \theta} \sigma^{(\varphi )} \partial_\varphi + {1 -
\alpha \over r} \sigma^{(r)} \right]\ ,
\ee
with $\sigma^{(r)} = \vec{\sigma} \cdot \hat{\vec{r}}\ ,\
\sigma^{(\theta)} = \vec{\sigma}\cdot \hat{\vec{\theta}}\ ,\
\sigma^{(\varphi )} = \vec{\sigma}\cdot \hat{\vec{\varphi } } $,
where $\hat{\vec r}\ ,\ \hat{\vec \theta}$ and $\hat{\vec
\varphi}$ are the standard unit vectors along the three spatial
directions in spherical coordinates. 

The Feynman two-component propagator obeys the equation 
\be
i\N {\cal S_F}(x,x') = {1 \over \sqrt{-g}} \delta^{(4)} (x - x')
I_2\ ,
\ee
and can be given in terms of the bispinor ${\cal G_F}$ by 
\be
{\cal S_F}(x,x') = i\N {\cal G_F}(x,x')\ , \label{Relation}
\ee
where now, ${\cal G_F}(x,x')$ obeys the $2\times 2$ differential
equation 
\be
\overline{{\cal L}} {\cal G_F}(x,x') = - {1\over \sqrt{-g}}
\delta^{(4)} (x - x') I_2\ ,
\ee
with
\be
\overline{\cal L} =-{1 \over \alpha^2} \partial^2_t + {\alpha^2
\over r^2} \partial_r r^2 \partial_r - {1\over r^2} \vec{L}^2
-{1-\alpha \over r^2}(1 + \vec{\sigma}\cdot \vec{L}) \ . 
\label{Operator}
\ee
The vacuum average value of the energy-momentum tensor may
be expressed in terms of the Euclidean Green function which is
simpler than the ordinary Feynman Green function. They are
connected by the relation \cite{BirrellDevies} ${\cal 
G_E}(\tau,\vec{r}; \tau', \vec{r'}) = -i {\cal G_F}(x,x')$,
where $t=i\tau$. In the following we shall consider the
Euclidean Green function. In order to find a solution for the
bispinor ${\cal G_E}(x,x')$, we shall obtain the solution for the
eigenvalue equation 
\be
\overline{\cal L} \phi_\lambda (x) = -\lambda^2 \phi_\lambda
(x)\ ,
\label{Eigenfunction}
\ee
with $\lambda^2 \geq 0$, so we can write 
\be
{\cal G_E}(x,x') = \sum_\lambda {\phi_\lambda (x) \phi_\lambda^+
(x') \over \lambda^2}\ . 
\ee 
Due to fact that our operator (\ref{Operator}) is self-adjoint,
the set of its eigenfunctions constitutes a basis for the Hilbert space
associated with two-component spinors. Moreover, because
operator $\overline{\cal L}$ is a parity even operator, its
eigenfunctions present a defined parity, so the normalized
eigenfunctions can be written as:
\bn
\phi_\lambda^{(1)}(x) &=& e^{-iE\tau}
f^{(1)}(r)\varphi^{(1)}_{j,m_j} (\theta ,\varphi )\ ,\\
\phi_\lambda^{(2)}(x) &=& e^{-iE\tau}
f^{(2)}(r)\varphi^{(2)}_{j,m_j} (\theta ,\varphi )\ , \nonumber
\en 
where $\varphi^{(k)}_{j,m_j}$, with $k=1,2$, are the spinor
spherical harmonics which are eigenfunctions of the operators
$\vec{L}^2$ and $\vec{\sigma}\cdot \vec{L}$ as shown below: 
\bn
\vec{L}^2\varphi^{(1,2)}_{j,m_j} &=& l(l+1)
\varphi^{(1,2)}_{j,m_j}\ ,\\
\vec{\sigma}\vec{L} \varphi^{(1,2)}_{j,m_j} &=& -(1 +
\kappa^{(1,2)}) \varphi^{(1,2)}_{j,m_j}\ , \label{Kappa}
\en 
with $\kappa^{(1)} = -(l+1) = -(j + 1/2)$ and $\kappa^{(2)} = l
= j + 1/2$. Explicit form of above standard function are given
in Ref. \cite{Bjorken}, for example.  

Substituting $\phi_\lambda^{(1,2)}$ into (\ref{Eigenfunction}),
we obtain the following eigenfunctions
\bn
\phi^{(k)}_\lambda(x) &=& \sqrt{p \over 2\pi r} e^{-iE\tau}
J_{\nu_k} (pr) \varphi^{(k)}_{j,m_j}\ , \label{Phi}\\
\lambda^2 &=& E^2/\alpha^2 + \alpha^2 p^2\ , \nonumber \\
\nu_1 &=& {l+1 \over \alpha} - {1\over 2}\ ,\ \nu_2 = {l\over
\alpha} + {1 \over 2}\ , \label{Nu} 
\en
where $J_\nu$ is the Bessel function of first kind. We can see
that for Minkowski spacetime where $\alpha = 1$ we have $\nu_1=
\nu_2 = l + 1/2$.

Now we are in condition to obtain the bispinor ${\cal G_E}$
which is given by: 
\be
{\cal G_E}(x,x') =
\int_{-\infty}^{+\infty} dE \int_0^\infty dp \sum_{j,m_j}
{\phi_\lambda^{(1)}(x)\phi_\lambda^{(1)+}(x') + 
\phi_\lambda^{(2)}(x)\phi_\lambda^{(2)+}(x') \over
E^2/\alpha^2 + \alpha^2 p^2}\ . \label{GE}
\ee  
Finally, substituting our result for the two-component spinor
$\phi_\lambda^{(k)}$ given by (\ref{Phi}) and (\ref{Nu}) into
(\ref{GE}) we obtain, with the help of Ref. \cite{Gradshtein},
an expression for the Euclidean Green function 
\be
{\cal G_E} (x,x') = {1 \over 2\pi r r'} \sum_{j,m_j} \left[
Q_{\nu_1 - 1/2} (u) C^{(1)}_{j,m_j}(\Omega ,\Omega') + Q_{\nu_2
- 1/2} (u) C^{(2)}_{j,m_j}(\Omega , \Omega')\right]\ , \label{GEsimple}
\ee
where $Q_\nu(u)$ are the Legendre functions of second kind; $u = 1
+ (\alpha^4 \triangle \tau^2 + \triangle r^2 )/2rr'$ and
$C^{(k)}_{j,m_j}(\Omega ,\Omega') = \varphi^{(k)}_{j,m_j}(\Omega
) \varphi^{(k)+}_{j,m_j}(\Omega')$.  
Again we can see that in Minkowski limit, for $\alpha =1$,
$\nu_1 = \nu_2 = l+1/2$ and we get
\be
{\cal G_E}(x,x') = {1 \over 8\pi^2}{1 \over \sigma_M(x,x')} I_2\
,
\ee
where we have used the relation envolving the sum of Legendre
functions and polynomials \cite{Gradshtein}. Here
$\sigma_M(x,x') = (\triangle\tau^2 + (\vec{r} - \vec{r}')^2)/2$
is one-half the square of geodesic distance between $x$ and $x'$
in the flat Euclidean space.  

Now in order to obtain the Green function ${\cal S_F}$ let us go
back to Eq. (\ref{GEsimple}). Using (\ref{Relation}) we have 
\be
{\cal S_F}(x,x') = i\left[{1 \over \alpha} \partial_t - \alpha
\sigma^{(r)}\partial_r + {1\over r}\sigma^{(r)} \vec{\sigma}
\cdot \vec{L} + {1- \alpha \over r} \sigma^{(r)} \right] {\cal
G_F}(x,x')\ .
\label{S_F}
\ee
\section{Vacuum Expectation Values}\label{EMT}
Now let us proceed with the calculation of the vacuum
expectation value (VEV) of the energy-momentum tensor.
Initially, we would like to discuss the general structure of
this tensor. As it will be seen later and has already been
discussed in scalar case in Ref. \cite{MazitelliLousto}, the
renormalized VEV of the energy-momentum tensor has the following
structure 
\be
<T^\nu_\mu>^{ren} = {1 \over 8\pi^2 r^4}\left[ A^\nu_\mu +
B^\nu_\mu \ln {\mu r\over \alpha}\right]\ , \label{Tmn}
\ee
where the tensors $A^\nu_\mu$ and $B^\nu_\mu$ depend on
the metric parameter $\alpha$, only. The scaling parameter $\mu$
appears after renormalization procedure. Obviously these tensors
are diagonal and the component $A^\theta_\theta =
A^\varphi_\varphi$ and $B^\theta_\theta = B^\varphi_\varphi$ due
to spherical symmetry of the problem. Therefore we have six unknown
components. The renormalized VEV of the energy-momentum tensor
must be conserved, i.e.,  
\be
<T^\nu_\mu>^{ren}_{;\nu} =0\ , \label{Conservation}
\ee
and gives the right conformal anomaly \cite{Wald77}, which for
massles spinor two-component field, reads \cite{ChristensenDuff} 
\be
<T^\mu_\mu>^{ren} = {1\over 16\pi^2} {\rm tr}a_2 = {T \over
8\pi^2 r^4}\ . \label{Trace}
\ee
Taking into account Eqs.(\ref{Conservation}), (\ref{Trace}) we may
express tensors $A^\nu_\mu$ and $B^\nu_\mu$ in terms of the
zero-zero components $A^0_0\ ,\ B^0_0$ and the trace $T$ by: 
\bn
A^\nu_\mu &=& {\rm diag}(A^0_0\ ;\ -T + A^0_0 + B^0_0\ ;\ T - A^0_0
-{1\over 2} B^0_0\ ;\ T - A^0_0 - {1 \over 2} B^0_0)\ , \\ 
B^\nu_\mu &=& B^0_0{\rm diag}(1;1;-1;-1)\ . \label{AnmBnm}
\en
Therefore our problem now is to obtain the zero-zero components
$A^0_0$ and $B^0_0$. Using the point-splitting approach, 
the VEV of the energy-momentum tensor for spinor field has the
following form (see Ref. \cite{BirrellDevies} for example): 
\be
<T_{\mu\nu}> = {1\over 4}\lim_{x' \to x} {\rm tr} \left[ \sigma_{\mu}
(\nabla_\nu - \nabla_{\nu'}) + \sigma_\nu (\nabla_\mu -
\nabla_{\mu'})\right] {\cal S_F}(x,x')\ ,
\ee
by means of which we have 
\be
<T^0_0> = {i \over \alpha^2} \lim_{x' \to x} \partial^2_t {\rm
tr}({\cal G_F}(x,x')) = -{1 \over \alpha^2} \lim_{x' \to x}
\partial^2_\tau {\rm tr}({\cal G_E}(x,x'
))\ . \label{T00}
\ee
The first term with time derivative in Eq. (\ref{S_F}) gives
non-zero contribution in the zero-zero component of the
energy-momentum tensor, only. Indeed, the Euclidean Green function
(\ref{GEsimple}) is proportional to the unit matrix $I_2$ after
taking the coincidence limit $\Omega' = \Omega$ and sum over
$m_j$.  The same is true for the third term in (\ref{S_F}) due to 
Eq. (\ref{Kappa}). We obtain zero contribution from these terms
because Pauli matrices are traceless. In above expression we have
used also that the Green function (\ref{GEsimple}) depends on the
$\triangle \tau$ and hence $\partial_{\tau'}{\cal G_E} = -
\partial_{\tau }{\cal G_E}$. 

Taking the coincidence limit $\Omega' =\Omega\ ,\ r'=r$ into 
Eq. (\ref{GEsimple}), summing over $m_j$ and after using the
integral representation for the Legendre function
\cite{Gradshtein}, it is possible to develop the sum over $j$,
which is geometric series and we arrive at the following 
formula for the Euclidean Green function 
\be
8\pi^2 {\cal G_E}(\triangle \tau ,r) ={1 \over 2r^2} \int_b^\infty
{dx \over \sqrt{x^2 -b^2}} {1 \over {\rm sh}^2 ({{\rm arsh} x \over
\alpha})}I_2\ , \label{GEt}
\ee
where the function $b$ may be expressed in terms of the one-half
of the square geodesic distance $\sigma$ in $\tau$ direction  
\be
b^2={\alpha^2 \over 2r^2} \sigma = {\alpha^2 \over 2r^2}\left( {
\alpha^2 \triangle \tau^2 \over 2}\right)\ .
\ee
In the case when only the angular variable $\Omega' =
\Omega$ coincides, we shall get the same formula for Green function 
(\ref{GEt}) with $\sigma = (\alpha^2 \triangle\tau^2 + \triangle
r^2/ \alpha^2)/2$ and $r^2 \to rr'$. 

The Green function (\ref{GEt}) is divergent in the coincidence
limit $b\to 0$. For renormalization in the massless case it is
more suitable to subtract from the Green function (\ref{GEt})
the Green function in the Hadamard form given below (see Refs.
\cite{ALNg,WaldPRD}) 
\be
8\pi^2{\cal G_E^H}(x,x') = \triangle^{1/2} \left\{ {a_0 \over
\sigma }  -{3a_2 \over 8}\sigma + \left[- {a_1 \over 2} + {a_2
\over 4} \sigma \right]\ln ({1\over 8}\mu^2 \sigma)
\right\}\ . 
\label{Hadamard}
\ee
In order to obey the conservation law (\ref{Conservation}) we
must subtract from the energy-momentum tensor additional contribution
$g_{\mu\nu} {\rm tr}a_2/64\pi^2$ according to Wald
\cite{WaldPRD}. The general form of the coefficients $a_k$ may
be found in Ref. \cite{Christensen78} and for the global monopole
spacetime they have the following form 
\be
\triangle = 1\ ,\ a_0=I_2\ ,\ a_1 = -{1 - \alpha^2 \over
6r^2}I_2\ ,\ a_2 = -{1- \alpha^4 \over 60r^4}I_2\ .  
\ee
From our expression for the Green function (\ref{GEt}), it is
possible to extract all divergencies in manifest form. To do
this let us consider the rhs of Eq. (\ref{GEt}). We divide the
integral in two parts - first one from $b$ to unit and second
from unit to infinity. In the first part we have the form below  
\be
V_1 = {1 \over 2r^2} \int_b^1 {dx \over \sqrt{x^2 - b^2}}{1
\over {\rm sh}^2({{\rm arsh} x \over \alpha})} I_2\ .
\ee
Subtracting and adding into the integrand the three first terms
of the power series of the function
\be
{1 \over {\rm sh}^2 ({{\rm arsh} x \over \alpha})} = {\alpha^2
\over x^2} - {1 - \alpha^2 \over 3} + {1 - \alpha^4 \over 15}{x^2
\over \alpha^2} + \dots\ , 
\label{Expansion}
\ee 
we get 
\be
V_1 = V_1^{fin}  + V_1^{div} \label{V_1}\ .
\ee
The first term is given by expression 
\be
V_1^{fin} = {1 \over 2r^2} \int_b^1 {dx \over \sqrt{x^2 -
b^2}}\left\{ {1 \over {\rm sh}^2({{\rm arsh} x \over \alpha})} - 
{\alpha^2 \over x^2} + {1 - \alpha^2 \over 3} - {1 - \alpha^4
\over 15}{x^2 \over \alpha^2}\right\}I_2\ ,
\label{V_1fin}
\ee
and its two derivatives with respect to time $\tau$ are finite
in the coincidence limit $b \to 0$. All divergencies are
contained in the second part of Eq. (\ref{V_1}) which has the form 
\be
V_1^{div} = {1 \over 2r^2} \left[ {\alpha^2 \over b^2}
\sqrt{1 - b^2} - {1 - \alpha^2 \over 3} \ln {1 + \sqrt{1 - b^2}
\over b} + {1 - \alpha^4 \over 30\alpha^2}\{ \sqrt{1 - b^2} +
b^2 \ln{1 + \sqrt{1- b^2}\over b}\} \right]I_2\ . \label{Exp}
\ee
Expanding this expression in the power series over $b$ 
up to the terms which will survive after taking the second
derivative and coincidence limit, we may represent it in the form
\be
V_1^{div} = \left\{ {1 \over \sigma} + {1 \over 2r^2}\left[ -{\alpha^2
\over 2} + {1 -\alpha^4 \over 30\alpha^2}\right] + 
\left[ -{\alpha^4 \over 32r^4} + {\alpha^2 (1 -\alpha^2) \over
48r^4} +  {a_2 \over 4}\right] \sigma + \left[ - {a_1
 \over 2} + {a_2 \over 4}\sigma 
\right] 
\ln {\alpha^2 \sigma
\over 8r^2} \right\} I_2\ .  
\ee
We see that the divergent part of this expression has the Hadamard
form given by Eq.(\ref{Hadamard}). The next powers in the
expansion (\ref{Expansion}) will give finite contribution to
the energy-momentum tensor. 

The second part 
\be
V_2 = {1 \over 2r^2} \int_1^\infty {dx \over \sqrt{x^2 - b^2}}{1
\over {\rm sh}^2({{\rm arsh} x \over \alpha})} I_2\ , 
\label{V_2}
\ee
does not contain divergencies and it may be easily expanded in
terms of $b$
\be
V_2 = {1 \over 2r^2} \int_1^\infty {dx \over x}{1
\over {\rm sh}^2({{\rm arsh} x \over \alpha})} I_2 + {\alpha^2
\sigma \over 8r^4}\int_1^\infty {dx \over x^3}{1 \over
{\rm sh}^2({{\rm arsh} x \over \alpha})} I_2 + O(\sigma^2)\ .   
\ee 
Taking into account all above formulas we obtain the following
expression for the renormalized Green function 
\bn
8\pi^2 {\cal G_E}^{ren} &=& 8\pi^2 [{\cal G_E}(\triangle \tau,
r) - {\cal G_E^H}(\triangle \tau, r)] \label{GEren}\\ 
&=& V_1^{fin} + V_2 + \left\{ {1 \over 2r^2}\left[
-{\alpha^2 \over 2} + {1 -\alpha^4 \over 30\alpha^2}\right] +
\left[ -{\alpha^4 \over 32r^4} + {\alpha^2 (1 -\alpha^2)
  \over 48r^4} + {5 a_2 \over 8}\right] \sigma
  + \left[  a_1 - {a_2 \over 2}\sigma 
\right] 
\ln {\mu r \over \alpha} \right\} I_2\ , \nonumber 
\en
where $V_1^{fin}$ and $V_2$ are given by eqs. (\ref{V_1fin}) and
(\ref{V_2}), respectively. Plugging this expression for
the renormalized Green function into Eq. (\ref{T00}) we obtain the
following formulas for the zero-zero components of $A^0_0$ and
$B^0_0$ (see Appendix \ref{A})  
\bn
A^0_0 &=& - {1 \over 8} \left\{ \int_0^1 {dx\over x} \left[
\left({\alpha^2 \over {\rm sh}^2({{\rm arsh} x \over
\alpha})}\right)'' - {6\alpha^4 \over x^4} - {2(1 - \alpha^4)
\over 15} \right] + \int_1^\infty {dx\over x} \left[
\left({\alpha^2 \over {\rm sh}^2({{\rm arsh} x \over
\alpha})}\right)'' - {6\alpha^4 \over x^4} \right] \right\}\
,\label{A00} \\ 
B^0_0 &=& {r^4{\rm tr} a_2 \over 2} = -{1 - \alpha^4 \over 60}\
, \label{B00}
\en
where the prime means the derivative with respect to $x$. In
this expression we have already taken into account additional
contribution $-{\rm tr}a_2/64\pi^2$. 

\begin{figure}[tbh]
  \centerline{
\psfig{figure=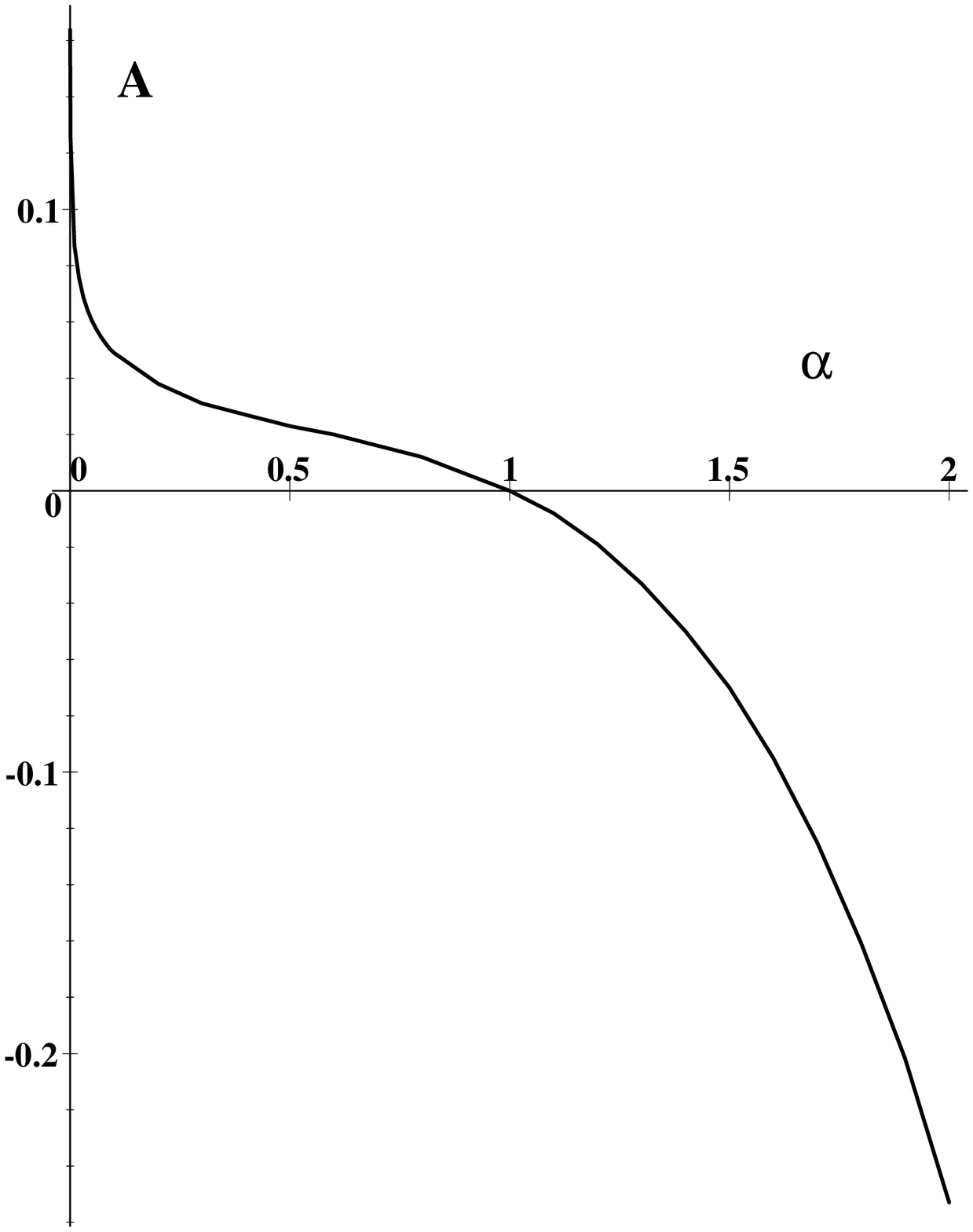,angle=0,height=8cm}%
\psfig{figure=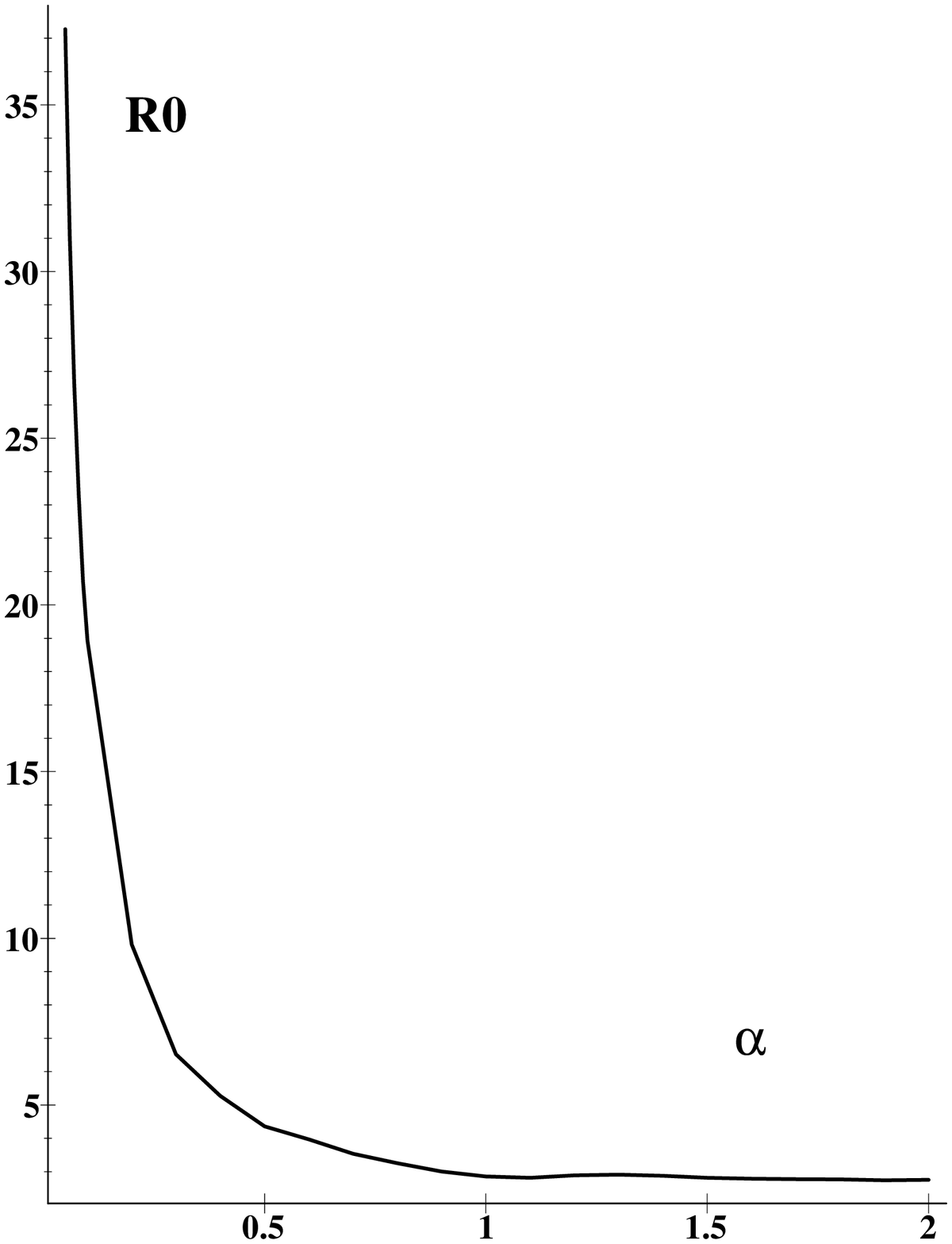,angle=0,height=8cm}}
\caption{Zero-zero component $A^0_0$ and distance $R_0$ where
the component $<T^0_0>^{ren}$ of energy-momentum tensor changes
its sign, are plotted as function of the metric parameter $\alpha$}
\label{1} 
\end{figure}

Now we have explicit expressions for zero-zero components $A^0_0$
and $B^0_0$ for arbitrary values of $\alpha$. Let us analyse 
$A^0_0$ component (see Appendix \ref{B} for details). 
\begin{enumerate}
\item For large solid angle deficit ($\alpha \ll 1$):  
\be
A^0_0 \sim -{1\over 60} \ln \alpha + C_0\ ,\ C_0 = 0.0104\ .
\ee
\item For small solid angle deficit (excess) ($|\alpha - 1|\ll
1$): 
\be
A^0_0 \sim C_1 (1 -\alpha)\ ,\ C_1 = 0.0773\ .
\ee  
\item The large solid angle excess ($\alpha \gg 1$):  
\be
A^0_0 \sim - C_\infty \alpha^4\ ,\ C_\infty = 0.0173\ .
\ee
\end{enumerate}
The numerical calculation of $A^0_0$ is shown in Fig.\ref{1} 

Now, we would like to compare our results with the
similar one for the VEV of the energy-momentum tensor of the
massless spinor field in the infinitely thin cosmic string
spacetime with line element, in cylindrical coordinate system, 
given by 
\be
ds^2 = -dt^2 + d\rho^2 + {\rho^2 \over \nu^2} d\varphi^2 +
dz^2\ . \label{InfThin}
\ee
In Ref. \cite{FrolovSerebr} has been obtained the following result  
\be
T^\nu_\mu = {(\nu^2 -1)(7\nu^2 + 17) \over 1440\cdot 4\pi^2
\rho^4} {\rm 
diag} (1;1;1;-3)\ .
\ee
In this case there is no the logarithmic contribution because
this spacetime is locally flat. Therefore we may compare now
the $A^0_0$ given by Eq. (\ref{A00}) for global monopole
spacetime with the analogous one for infinitely thin cosmic
string spacetime which is given by expression
\be
A^0_{0(cs)} = {(\nu^2 -1)(7\nu^2 + 17) \over 720}\ .
\ee 
To do so, we have to change the radial variable $r = \alpha \rho$
because in this coordinate system the section $\theta = \pi /2$
of the monopole spacetime (\ref{Metric}) coincides with section
$z=const$ of the infinitely thin cosmic string spacetime
(\ref{InfThin}) and the parameter $\alpha = 1/\nu$. Therefore we
may compare $A^0_{0(cs)}(\nu)$ with $(A^0_{0(gm)}(\alpha
)/\alpha^4)_{\alpha = 1/ \nu}$. We have qualitative agreement of
both these quantities. For $\nu \ll 1 $ they go to negative
constant; for small angle deficit they are proportional to $\nu -1$;
for $\nu \gg 1$ we have $A^0_{0(cs)} \sim \nu^4$ and 
$[\nu^4A^0_{0(gm)}(1/\nu)] \sim \nu^4 [\ln \nu + 60C_0]$. Both
of these quantities change the sign at point $\nu = 1$. 

The dependence of $<T^0_0>^{ren}$ on the distance from origin in
the global monopole spacetime is competely different from the
case of the infinitely thin cosmic string due to the logarithmic
term. Let us consider the physical distance $R=\mu r/\alpha$
measured in a mass scale $\mu$. Then, the $<T^0_0>^{ren}$
measured in units $\mu^4$ has the form 
\be
8\pi^2 \mu^{-4}<T^0_0>^{ren} = {\alpha^4 - 1 \over 60\alpha^4
R^4} \ln {R \over R_0}\ ,
\ee
where $R_0 = \exp (-A^0_0/B^0_0)$. In the case $\alpha < 1$ the
energy density $<T^0_0>^{ren}$ is positive in the domain $R\in (0,R_0)$;
it changes the sign in the point $R=R_0$ and goes, through
the minimum at point $R_* = R_0e^{1/4}$, to zero in infinity.
In the case $\alpha > 1$ we have opposite picture with maximum
at the point $R_*$. The dependence of $R_0$ on the $\alpha$
is shown in the Fig.\ref{1}. For small $\alpha$ it is proportional
to $1/\alpha$ and it goes to constant for great $\alpha$. 
\section{Conclusion}\label{Conclusion}
In this paper we have considered the quantum, a $1/2$-spin
left-handed field in the background of a pointlike global
monopole described by the metric tensor given in (\ref{Metric}).
More specifically, we have obtained the complete Feynman
propagator, expressed in terms of a bispinor, in a closed form.
Differently from the results obtained in Ref.
\cite{MazitelliLousto} for scalar field, in this analysis there 
appears two effective angular total quantum numbers, which we
call by $\nu_1$ and $\nu_2$. These angular total quantum numbers
are related with the explicit properties of the spinor
harmonics, $\varphi^{(1)}_{j,m_j}$ and $\varphi^{(2)}_{j,m_j}$
and they have the following form $\nu_1 = (l+1)/\alpha -1/2$ and
$\nu_2 = l/\alpha + 1/2$. For the scalar case the effective
orbital quantum numbers take a simple form only under specific
situation \cite{MazitelliLousto}, namely for $\xi = 1/8$.   

The main goal of this work was to obtain the renormalized
vacuum expectation value of the energy-momentum tensor. Because
all components of the VEV of the energy-momentum tensor can be
expressed in terms of its zero-zero component, we used only the
Euclidean Green function at the coincidence limit $\Omega' =
\Omega$ and $r' = r$. In this case, in order to obtain the
regularized expression for $<T^0_0>$, we were able to subtract
all divergencies in a manifest form.  

After the renormalization, the Green function (\ref{GEren}) depends
on the scale parameter $\mu$ which leads to logarithm
contribution to the energy-momentum tensor given by Eq.(\ref{Tmn}).
But, as it was noted in Ref. \cite{MazitelliLousto} the one-loop
Einstein equations 
\be
R_{\mu\nu} - {1\over 2}R g_{\mu\nu} +\epsilon_1 {}^{(1)}H_{\mu\nu} +
\epsilon_2 {}^{(2)}H_{\mu\nu} = 8\pi \{T_{\mu\nu}^{clas} + <T_{\mu
\nu}>^{ren}\}\ ,
\ee
do not depend on this parameter due to the renormalization group
equations for the coefficients $\epsilon_1$ and $\epsilon_2$. Any
variation of the scale parameter $\mu$ may be absorbed by
variation of the $\epsilon_k$. 

Taking into account the conservation law (\ref{Conservation}), the
expression for conformal anomaly (\ref{Trace}) and the spherical
symmetry of problem we expressed all components of
energy-momentum tensor in terms of zero-zero component
(\ref{AnmBnm}) which has the form
\be
<T^0_0>^{ren} = {1 \over 8\pi^2 r^4}\left[ A^0_0 +
B^0_0 \ln {\mu r\over \alpha}\right]\ , \label{ApT00}
\ee 
where $A^0_0$ and $B^0_0$ are given by eqs. (\ref{A00}) and
(\ref{B00}). The component $A^0_0$ depicted in Fig.1 as a
function of $\alpha$, and it is qualitatively agreed with
similar one in the infinitely thin cosmic string spacetime. 

The scaling parameter $\mu$ leads to the logarithmic
contribution in the energy-momentum tensor (\ref{ApT00}). For
this reason $<T^0_0>^{ren}$ changes its sign in some point $R_0$ and has
the extremum at the point $R_* = R_0 e^{1/4}$. The dependence of
$R_0$ on the $\alpha$ is depicted in the Fig. \ref{1}. 

Before we finish this paper we would like to make a brief
comment about some results previously obtained in the literature
related with quantum calculation of the energy-momentum tensor
$T^\nu_\mu$ in the spacetime of a pointlike global monopole.
Hiscock, in Ref. \cite{Hiscock}, using general consideration,
obtained a formal expression for the vacuum expectation value of
the energy-momentum tensor for an arbitrary collection of
conformal massless fields in this manifold. Later, Mazzitelly
and Lousto, in Ref. \cite{MazitelliLousto}, developed the
explicit calculation for renormalized vacuum average of the
square of a massless scalar quantum field and also, by general
consideration, infered the structure for $T^\nu_\mu$ which
disagrees with the Hiscock's results and manifestly depends on
the scale parameter $\mu$. This is in agreement with general
consideration by Wald in Ref. \cite{WaldPRD}. In our paper we
obtained the explicit expression for vacuum average value of the
energy-momentum tensor for a massless fermionic field in this
manifold, giving the complete information about this tensor in
terms of $A^0_0$ and $B^0_0$. We also have analyzed the
behaviour of these functions with the parameter $\alpha$, and
the dependence of $T^0_0$ with the physical distance $R$.
Finally we want to say that our expression for VEV of the
energy-momentum tensor depend on the scale parameter $\mu$, in
agreement with Ref. \cite{WaldPRD}, due to the appearance of
a logarithm term, which is a consequence of the regularization
procedure. This term is responsible for the changing in the sign
of the vacuum energy-momentum tensor when we vary the physical
distance $R$. 

The explicit expression for the massless left-handed
two-component spinor Green function, is obtained in the
background of global monopole spacetime. It is our interest to
develop a similar calculation for the massive case and also
obtain the vacuum average for the energy-momentum tensor.
\section*{Acknowledgments}
NK is grateful to Departamento de F\'{\i}sica, Universidade
Federal da Paraiba (Brazil) where this work was done, for
hospitality. His work was supported in part by CAPES and in part
by the Russian Fund for Basic Research, grant No 97-02-16318.  

ERBM and VBB also would like to thank the Conselho Nacional de
Desenvolvimento Cientifico e Tecnol\'ogico (CNPq). 
\appendix
\section{}\label{A}
In order to find derivatives of $V_1^{fin}$ with respect to $b$
we have to change variable $x \to b x$ because integrand in
$V_1^{fin}$ is divergent at the point $x= b$. The second
derivative has the form below 
\be
2r^2{d^2 V_1^{fin} \over d b^2} = {1 - b^2 \over b^2(1
- b^2)^{3/2}} f(1) - {1 \over b^2 \sqrt{1 - b^2}} f'(1) + {1
\over b^2} \int_b^1 {dx x^2 \over \sqrt{x^2 - b^2}} f''(x)\ , 
\ee
where
\be
f(x) = {1 \over {\rm sh}^2({{\rm arsh} x \over \alpha})} - 
{\alpha^2 \over x^2} + {1 - \alpha^2 \over 3} - {1 - \alpha^4
\over 15}{x^2 \over \alpha^2}\ .
\ee
The last integral may be represented in the following form
\be
{1 \over b^2 }\int_b^1 {dx x^2 \over \sqrt{x^2 - b^2}}f''(x) =
{\sqrt{1 - b^2} \over b^2}(f'(1) - f(1)) + \int_b^1 {dx \over
\sqrt{x^2 - b^2}} \left[f''(x) - {1\over x} f'(x) + {1 \over
x^2} f(x)\right]\ .
\ee
Taking into account the above expression and putting $b=0$ we have
that 
\be
2r^2{d^2 V_1^{fin} \over d b^2}_{b=0} = 
- f'(1) +\int_0^1 {dx \over x} 
      \left[f''(x) - {1\over x} f'(x) + {1 \over x^2} f(x)
      \right]\ .
\ee
Integrating by parts the term with second derivative we
arrive at the expression
\be
{d^2 V_1^{fin} \over d b^2}_{b=0} = {1\over 2r^2} 
\int_0^1 {dx \over x^3} f(x)\ . 
\ee
The second derivative of $V_2$ may be easily found from Eq.
(\ref{V_2}) and is given by
\be
{d^2 V_2 \over d b^2}_{b=0} = {1\over 2r^2} 
\int_1^\infty {dx \over x^3} {1 \over {\rm sh}^2({{\rm arsh} x \over
\alpha})}\ .  
\ee
Using the above formulas in (\ref{GEren}) and (\ref{T00}), we get
the following expression for $<T^0_0>^{ren}$ 
\bn
8\pi^2<T^0_0>^{ren}=&& -{\alpha^2 \over 4 r^4}\left\{
\int_0^1 {dx \over x^3}\left[ {1 \over {\rm sh}^2({{\rm arsh} x
\over \alpha})} - {\alpha^2 \over x^2} + {1 - \alpha^2 \over 3}
- {1 - \alpha^4 \over 15}{x^2 \over \alpha^2}\right] \right. \\ 
&&+\left. \int_1^\infty {dx \over x^3}\left[ {1 \over {\rm
sh}^2({{\rm arsh} x \over \alpha})} - {\alpha^2 \over x^2} + {1
- \alpha^2 \over 3}\right] \right\} + {{\rm tr}a_2 \over 2}
\ln{\mu r \over \alpha} - {5 \over 8}{\rm tr}a_2 - {1 \over 8}
{\rm tr}a_2\ .  \label{A8}
\en
Because we have used for renormalization of the Green function in
the Hadamard form (\ref{Hadamard}) the last term has been added in
order to obey conservation law (\ref{Conservation}) according to
Wald \cite{WaldPRD} . Integrating two times by parts we arrive at
the formula (\ref{A00}); the last two terms in (\ref{A8}) are
cancelled.  
\section{}\label{B}
Here we will analyse $A^0_0$ component in three domains :
$\alpha \ll 1\ ,\ |\alpha - 1| \ll 1$ and $\alpha \gg 1$.
$A^0_0$ has the form below
\be
A^0_0 = -{1 \over 8} \left\{ \int_0^1 {dx \over x} f_2 +
\int_1^\infty {dx \over x} f_1\right\}\ ,
\ee
where 
\bn
f_1 &=& {2\alpha x\over (1 + x^2)^{3/2}} {{\rm ch} y \over {\rm
sh}^3 y} - {2 \over 1+x^2} \left( {1 \over {\rm sh}^2 y } -
{3{\rm ch}^2 y \over {\rm sh}^4 y}\right) - 6{\alpha^4 \over
x^4}\ ,\nonumber \\ 
f_2 &=& f_1 -{2 \over 15}(1 - \alpha^4)\ ,\ y={1\over
\alpha}{\rm arsh} x \ .
\en

\noindent
1. Case $\alpha \ll 1$. \\
Here we represent $A^0_0$ in the following form 
\be
A^0_0 = -{1 \over 8}\left\{ \int_0^\alpha {dx\over x} f_3 +
\int_\alpha^\infty{dx \over x} f_2 - {2 \over 15} (1 - \alpha^4)
\int_\alpha^1 {dx\over x}\right\}\ . 
\ee
The first and second integrals are finite in the limit $\alpha
\to 0$ (after changing $x\to \alpha x$) and the last integral gives
logarithm term. Putting together we obtain
\bn
A^0_0 &=& -{1 \over 60}\ln \alpha + C_0\ ,\\
C_0   &=& {1\over 4} \left\{ \int_0^1 {dx \over x} \left( {1 \over
{\rm sh}^2 x} - {3{\rm ch}^2 x\over {\rm sh}^4x} + {3 \over x^4} + {1
\over 15}\right) + \int_1^\infty {dx \over x} \left( {1 \over
{\rm sh}^2 x} - {3{\rm ch}^2 x\over {\rm sh}^4x} + {3 \over x^4
}\right) \right\} = 0.0104\ .
\en

\noindent
2. Case $|\alpha - 1| \ll 1$. 
There is no problem to expand $A^0_0$ near the point $\alpha
= 1$ and we get:
\be
A^0_0 = C_1 (1 - \alpha)\ ,\ C_1 = {7 \over 900} + {1 \over 15}
\ln 2 = 0.0773
\ee  

\noindent
3. Case $\alpha \gg 1$. 
In this case 
\bn
A^0_0 =&& -C_\infty \alpha^4\ , \\
C_\infty = && {1\over 8} \left\{ \int_0^1 {dx \over x} \left( {2x
\over (1 + x^2)^{3/2}} {1 \over {\rm arsh}^3 x} + {6 \over 1 +
x^2} {1 \over {\rm arsh}^4 x} - {6 \over x^4} + {2\over 15}
\right) \right. \nonumber \\
&& +\left. \int_1^\infty {dx \over x} \left( {2x
\over (1 + x^2)^{3/2}} {1 \over {\rm arsh}^3 x} + {6 \over 1 +
x^2} {1 \over {\rm arsh}^4 x} - {6 \over x^4} \right)\right\} =
0.0173 \nonumber
\en

\end{document}